\documentclass[12pt]{article}
\usepackage{lmodern} % safe main font
\usepackage{fontspec}
\usepackage{xeCJK}
\usepackage[margin=1in]{geometry}
\usepackage{amsmath}
\usepackage{graphicx}
\usepackage{tikz-cd}
\usepackage{multicol}
\usepackage{booktabs}
\usepackage{tabularx}
\usepackage{array}
\usepackage{multirow}
\usepackage{setspace}
\usepackage{fancyhdr}
\usepackage{ulem}
\usepackage[hidelinks]{hyperref}
\usepackage[backend=biber,style=apa,sorting=nyt]{biblatex}
\DeclareLanguageMapping{english}{english-apa}
\AtBeginBibliography{\raggedright}

\makeatletter
\renewcommand{\section}{\@startsection{section}{1}{\z@}%
  {-3.5ex \@plus -1ex \@minus -.2ex}%
  {2.3ex \@plus.2ex}%
  {\centering\normalfont\Large\bfseries}}
\renewcommand{\subsection}{\@startsection{subsection}{2}{\z@}%
  {-3.25ex\@plus -1ex \@minus -.2ex}%
  {1.5ex \@plus .2ex}%
  {\centering\normalfont\large\bfseries}}
\renewcommand{\subsubsection}{\@startsection{subsubsection}{3}{\z@}%
  {-3.25ex\@plus -1ex \@minus -.2ex}%
  {1.5ex \@plus .2ex}%
  {\centering\normalfont\normalsize\bfseries}}
\makeatother

\begin{document}

\title{Restoring Heterogeneity in LLM-based Social Simulation: An Audience Segmentation Approach}
\author{%
  Xiaoyou Qin\\
  \small School of Journalism, Fudan University
  \and
  Zhihong Li\\
  \small College of Media and International Culture, Zhejiang University
  \and
  Xiaoxiao Cheng\thanks{Corresponding author: \texttt{xxcheng21@zju.edu.cn}}\\
  \small College of Media and International Culture, Zhejiang University
}
\maketitle

\clearpage

\section*{Abstract}

Large language models (LLMs) are increasingly used to simulate social attitudes and behaviors, offering scalable ``silicon samples'' that can approximate human data. However, current simulation practice often collapses diversity into an ``average persona,'' masking subgroup variation that is central to social reality. This study introduces audience segmentation as a systematic approach to restoring heterogeneity in LLM-based social simulation. Using US climate-opinion survey data, we compare six segmentation configurations across two open-weight LLMs (Llama 3.1-70B and Mixtral 8x22B), varying segmentation identifier granularity, parsimony, and selection logic (theory-driven, data-driven, and instrument-based). We evaluated simulation performance by using a three-dimensional evaluation framework covering distributional, structural, and predictive fidelity. The results show that increasing identifier granularity does not produce consistent improvement: Moderate enrichment can improve performance, but further expansion does not reliably help and can worsen structural and predictive fidelity. Across parsimony comparisons, compact configurations often match or outperform more comprehensive alternatives, especially in structural and predictive fidelity, while distributional fidelity remains metric dependent. Identifier selection logic determines which fidelity dimension benefits most: Instrument-based selection best preserves distributional shape, whereas data-driven selection best recovers between-group structure and identifier--outcome associations. Overall, no single configuration dominates all dimensions, and performance gains in one dimension can coincide with losses in another. In addition, all LLMs showed residual overregularization, indicating persistent difficulty in fully recovering real-world population heterogeneity. These findings position audience segmentation as a core methodological approach to valid LLM-based social simulation and highlight the need for heterogeneity-aware evaluation and variance-preserving modeling strategies.

\begin{center}
\textbf{Keywords:} LLM-based social simulation; silicon sample; segmentation; fidelity; heterogeneity
\end{center}

\clearpage

\section{Introduction}
Large language models (LLMs) have rapidly transformed social science research, creating new possibilities and capabilities for simulating human attitudes \parencite{Argyle2023OutOfOneMany,Lee2024EstimatePublicOpinion} and modeling complex social dynamics \parencite{Park2023GenerativeAgents,Gao2024LLMEmpoweredABM}. Their applications now extend far beyond conventional tasks such as literature review, text classification, and survey design \parencite{Jansen2023EmployingLLMs}. Of particular note is the emergent field of LLM-based social simulation---a promising yet methodologically demanding domain.

A pivotal development in this field is \textcite{Argyle2023OutOfOneMany}'s ``silicon sampling'' paradigm, which leverages LLMs to generate synthetic data capable of substituting for traditional human respondents. This brings numerous advantages, including dramatic reductions in research cost and time, as well as the ability to design and implement experiments that were previously infeasible due to practical or ethical constraints \parencite{Jansen2023EmployingLLMs,Rossi2024ProblemsLLMData}. In this line of research, LLM-based simulation largely falls into two main categories. The first integrates LLMs as agents within multi-agent systems to emulate complex decision-making processes, emergent collective behavior, and the corresponding social interaction patterns \parencite{Park2023GenerativeAgents,Tornberg2023SimulatingSocialMedia,Wu2025Boundary}. The second employs LLMs as tools for measuring public opinion and directly replicating, estimating, and predicting public attitudes on issues such as climate change, political preferences, and policy support \parencite{Lee2024EstimatePublicOpinion,Yang2024SocialPredictors}.

Nevertheless, while these advances are exciting and promising, they also bring into sharp relief a fundamental challenge: preserving heterogeneity. The performance of LLM-based simulation hinges on several widely recognized factors, such as training data quality, model transparency, prompt engineering, and benchmark selection \parencite{Argyle2023OutOfOneMany,Grossmann2023AITransformation,Dillion2023CanAIReplace}. However, a more insidious problem remains. LLM architectures, by design, tend to compress social diversity into their most probable---or modal---representations \parencite{Wang2025MisportrayFlatten}. This ``average persona'' effect \parencite{Wu2025Boundary,Rossi2024ProblemsLLMData}, wherein model responses converge toward central tendencies, largely obscures and even erases the very differences at the subgroup level that are essential for rich, meaningful, and credible social simulation. This phenomenon is not just a technical curiosity; it directly undermines attempts to accurately capture the distributional reality of social-grouping attitudes. Genuine social simulation, as \textcite{Lyman2025BalancingAlignment} warn, requires models that can both distinguish and generate responses from multiple subgroups, rather than models that reduce variation and heterogeneity to undifferentiated averages and homogeneity.

The erasure of heterogeneity, which refer to as ``heterogeneity masking,'' is multifaceted. While the technical compression partially stems from LLMs' inherent training mechanisms that favor averaged representations \parencite{Wang2025MisportrayFlatten}, the absence of heterogeneity thinking in simulation practices exacerbates the issue. Methodologically, prompt engineering efforts tend to focus on refining instruction clarity and persona consistency yet rarely addresses how to systematically encode population heterogeneity in simulations \parencite{Davidson2025IntegratingGenAI}. This is largely due to a lack of robust frameworks for translating group variance into research design. Epistemologically, there is often an implicit bias toward generalizability, in which ``representative'' means ``average'', rather than accurate reproductions of real-world diversity and distributional richness. 

Despite growing interest in demographic profiling and persona conditioning, systematic guidance on how segmentation choices affect simulation performance remains limited. Prior research has typically examined whether demographic characteristics should be included \parencites{Argyle2023OutOfOneMany,Santurkar2023WhoseOpinions}, but has paid less attention to which variables (hereafter termed ``segmentation identifiers'') matter, the level of granularity required, and the selection logic used. As a result, existing social simulation studies risk privileging modal responses and obscuring meaningful differences---often unintentionally.

This is where our study intervenes. We advocate for audience segmentation, an approach that has been well established within communication science and marketing research \parencite{Slater1995AmericanHealthstyles,Slater1996TheoryMethod} as a systematic way to restore heterogeneity in LLM-based simulation. Audience segmentation partitions populations into subgroups that are theoretically and empirically meaningful \parencite{Hine2014AudienceSegmentation}. This study employs climate opinion survey data collected from the United States to rigorously examine how different segmentation strategies affect LLMs' capacity to replicate human attitude distributions while preserving subgroup differences. 

To rigorously evaluate this proposition, we assessed six segmentation configurations that varied along the dimensions of identifier granularity, parsimony, and selection logic, and evaluated simulation performance using distributional, structural, and predictive fidelity. Our empirical analysis challenges the intuitive assumption that increasing the detail of persona prompts inherently yields more realistic social variation. Instead of improving fidelity, the indiscriminate accumulation of segmentation identifiers paradoxically amplifies the model's tendency toward over-regularization, which distorts the authentic relational structures among subgroups. We demonstrate that the successful restoration of heterogeneity depends fundamentally on informative parsimony. Furthermore, the underlying logic for identifier selection---whether theory-driven, data-driven, or based on empirical instruments—directly determines which specific dimension of simulation performance is most effectively preserved. Ultimately, these findings reposition audience segmentation from a surface-level prompt-engineering tactic into a core methodological decision for valid LLM-based research.

% The remainder of this paper proceeds as follows. Section 2 develops our theoretical framework distinguishing population thinking from typological thinking. Section 3 diagnoses how current practices mask heterogeneity at technical, methodological, and epistemological levels. Section 4 positions audience segmentation as the operationalization of variability principle. Section 5 lays out our experimental design and validation metrics. Section 6 presents the main findings. Section 7 concludes with practical recommendations and directions for future research.

\section{From Typological Thinking to Population Thinking}

The masking of heterogeneity epitomizes a deeper epistemological divide in the social sciences concerning how scholars conceptualize social variations. This divide is rooted in two philosophical traditions that have long shaped scientific inquiry: typological thinking and population thinking.

Typological thinking, rooted in the Platonic tradition, treats variation as unwanted noise around an essential ``type.'' Deviations from the type are regarded as confounding factors or exogenous disturbances that need to be eliminated in the pursuit of universal laws \parencite{Xie2013PopulationHeterogeneity}. A key assumption underlying this tradition, which has worked well in natural science, is homogeneity: Once we understand a type of phenomenon, we can generalize that knowledge to individual cases. Within the social sciences, typological thinking often manifests in attempts to identify modal personalities, average voters, or singular ``representative'' agents. Individual differences are interpreted as error variance that obscures the underlying ``true'' pattern. At its very core, this ``social physical'' way of thinking naively essentializes population averages, making the ``average man'' the primary object of exploration. While appealing for its simplicity and promise of generalizability, this approach is limited in terms of modeling social reality, which is characterized by structured variation across social groups.

In contrast, population thinking emerged from the Darwinian revolution and was later introduced to the social sciences by Francis Galton. In this mode of thinking, deviations from the mean are not scientifically trivial; they constitute the very basis of evolution and are, therefore, an intrinsic property of distributions rather than departures from an idealized type \parencite{Xie2013PopulationHeterogeneity,Xie2024LocalizationSociology}. Instead of privileging ``typical'' cases, population thinking treats heterogeneity---including subgroups, outliers, and the full spectrum of individual responses---as meaningful signals that reveal the underlying structure of social phenomena. Accordingly, heterogeneity, not homogeneity, becomes central to explanations of social change. 

This distinction is far from philosophical; it carries direct methodological implications for social simulation. Population heterogeneity, that is, the authentic diversity of attitudes, identities, and behaviors, constitutes a precondition for valid social simulation. Research on complex systems has established that agent diversity drives behavioral evolution and enables emergent phenomena that homogeneous populations cannot generate \parencite{Wu2025Boundary}.

Recent work has further underlined the relevance of population thinking for LLM-generated data. \textcite{Argyle2023OutOfOneMany}, for instance, demonstrated that demographically conditioned prompts can approximate survey responses for distinct subgroups. However, such attempts often assume, without sufficiently testing, whether conditioning truly preserves meaningful heterogeneity or merely produces tailored ``types'' or ``average men'' that still fail to capture the breadth of human diversity \parencite{Lyman2025BalancingAlignment}. 

This points to the central problem in LLM-based social simulation: Despite the rhetorical embrace of population-level accuracy, current practices often inadvertently revert to typological logic. Unconditioned or ill-specified prompts tend to yield statistically averaged, ``typical'' responses---types that erase the distributional richness central to social scientific inquiry. What is at stake, then, is not merely a technical limitation but an epistemological misalignment; to be specific, methods implicitly built on typological premises are at odds with the population-thinking ethos that underlies many sociological theories. Without deliberate intervention, simulations risk perpetuating and (re)producing the very ``average persona'' effect that undermines meaningful group-level and societal inference.

Addressing this misalignment requires more than technical tuning; it calls for population thinking that is operationalized and baking it into simulation design---specifically, through systematic audience segmentation strategies. Treating heterogeneity as a signal to be detected, rather than as noise to be minimized, not only aligns simulations with complex real-world realities but also enables more nuanced inquiry into mechanisms, subgroup dynamics, and social structure. 

\section{Sources of Heterogeneity Masking}
As outlined above, heterogeneity masking is a crucial yet understudied limitation in current LLM-based social simulation. Without deliberate methodological attention, simulated populations risk being reduced to overly simplistic ``average personas.'' This flattening of diversity does not occur by chance; it is shaped by technical, methodological, and epistemological choices made throughout the simulation process. Recognizing these mechanisms is a necessary first step toward a population-thinking approach.

\subsection{Technical compression}
Heterogeneity is often erased by mechanism embedded in LLM training and optimization. Mainstream LLMs are typically trained with methods such as maximum likelihood estimation (MLE), which incentivizes models to produce the most probable---and thus often the most central---response to a given prompt; this means those less-typical, minority, or outlier responses are naturally suppressed, thereby making LLMs struggle with preserving Significant population heterogeneity \parencite{Kirk2024PersonalizingAlignment}. Specifically, likelihood-based loss functions encourage high-probability generations, which compress diversity across subgroups \parencite{Wang2025MisportrayFlatten}. Reinforcement learning from human feedback can exacerbate this tendency: by aligning model outputs with what human annotators judge as ``high quality'' or the ``right answer,'' LLMs are nudged toward consensual, moderate, and ``safe'' responses, obscuring authentic social variation across social groups. \textcite{Santurkar2023WhoseOpinions} describe this process as ``value alignment,'' which prioritizes certain value systems and systematically reduces the expression of diverse viewpoints. 

The empirical consequences of these mechanisms have been documented in recent work. \textcite{Boelaert2025MachineBias} tested LLMs’ capacity to replicate survey responses across 687 sociodemographically defined subpopulations; even with explicit sociodemographic prompts, the resulting answer distributions remained highly concentrated across subgroups. Importantly, their analysis demonstrated that this prediction error cannot be attributed to a single, stable pattern of social bias; instead, errors shifted across groups and survey questions, implying that heterogeneity compression is not merely inherited from societal inequalities in training corpora; rather, it arises from representational compression in model architectures.

Taken together, these technical mechanisms create a foundational layer of heterogeneity masking that operates regardless of the researchers' intentions.

\subsection{Methodological neglect}
Methodological choices can compound technical compression when heterogeneity is not operationalized systematically. While proper conditioning is widely recognized as essential for simulation fidelity \parencite{Argyle2023OutOfOneMany}, prompt engineering is often treated as technical tuning rather than as a process with profound implications for data validity \parencite{Rossi2024ProblemsLLMData}. This produces ``ad hoc segmentation'' simulation practices that include identity cues without a clear theoretical rationale.

Some studies exhibit unreflective methodological inertia toward demographic conditioning, routinely including age, gender, education, and race without considering the domain dependence of segmentation identifiers or specific research contexts. However, key variables affecting simulation performance vary by research topic Indiscriminately piling demographic information does not necessarily improve fidelity and may introduce irrelevant noise \parencite{Anthis2025Position}. Recent empirical work offers a cautionary illustration: Despite conditioning prompts on five standard sociodemographic variables, LLMs failed to generate meaningfully differentiated responses across subgroups \parencite{Boelaert2025MachineBias}. This suggests that surface-level demographic characteristics function as ``identity labels'' rather than ``attitudinal anchors''; they inform the model of who someone is, but not how that person is likely to think or respond within a specific domain. More fundamentally, social identities are not merely individual attributes but expressions of group relations embedded in social structures \parencite{Turner1979SocialComparison}. Treating prompt variables as independent individual labels ignores in-group/out-group biases \parencite{Hu2025IdentityBiases} and strips group identity of its meaning within specific social contexts.

\subsection{Epistemological misalignment}

Perhaps the most insidious layer of heterogeneity masking is the epistemological layer. This manifests in two ways. First, researchers often pursue model generalizability and representativeness in ways that implicitly privilege averages. Models are expected to produce ``typical'' opinions rather than capture distributional variation. This carries over typological thinking-based sampling logic that is misaligned with simulation goals that require the preservation of social variations. Second, prevailing evaluation practices reinforce this bias. Many studies validate simulation fidelity using aggregated summary statistics while sidelining granular subgroup-level comparisons. Success is declared when model outputs ``match'' real-world distributions on overall means or medians, regardless of how well subgroup variation is preserved. 

This evaluative blind spot has tangible costs. \textcite{Wang2025MisportrayFlatten} compared LLM outputs against 3,200 human participants across 16 demographic identities and found that LLMs consistently ``flatten'' groups. Their responses occupy a far narrower semantic space than those of actual humans. Across four models and multiple diversity metrics, this pattern held for nearly every demographic group tested. More troubling still, LLM personas often resemble out-group imitations rather than authentic in-group voices, particularly for marginalized communities such as non-binary individuals and people with disabilities. The risk, then, is a drift toward superficial prediction rather than explanatory depth \parencite{Hofman2017PredictionExplanation}. Social mechanisms, group dynamics, and causal processes are drowned in the ``noise'' of averages \parencite{Hedstrom2010CausalMechanisms}.

\section{Audience Segmentation as a Pathway to Restore Heterogeneity}

The preceding section reveals that heterogeneity masking operates through interlocking technical, methodological, and epistemological mechanisms. Addressing the ``average persona'' problem therefore requires more than technical refinement; it demands an integrated and systematic framework that embeds population thinking into simulation design from the outset.

Here, we propose audience segmentation as such a framework. A clarification is necessary at the outset: The heterogeneity we seek to restore is not an exact replication of real-world population distributions, which would be neither feasible nor, for many research purposes, necessary. Rather, our goal is to recover theoretically meaningful segments---subgroups that are conceptually grounded and empirically distinguishable according to established social scientific frameworks. Preserving such theoretical segmentation is essential for understanding group-level mechanisms and dynamics, even if perfect demographic fidelity remains unattainable. Segmentation, in this sense, is not merely a technical procedure but the explicit operationalization of population thinking. It treats heterogeneity and variation as signals rather than noise, thereby realigning simulation practices with the realities of differentiated societies.

\subsection{Segmentation analysis}
Segmentation analysis originated in audience-oriented communication research and marketing science, where it serves to identify structural differences across psychological, behavioral, and value dimensions, thereby partitioning audiences into subgroups that are internally cohesive yet externally differentiated \parencite{Slater1995AmericanHealthstyles,Slater1996TheoryMethod}. Rather than arbitrary labeling, this method is a systematic effort to remain faithful to the diversity that lready exists in social structures. It recognizes that individual behavior is jointly shaped by social environments, perceived norms, and cognitive processes \parencite{Hine2014AudienceSegmentation}.

The epistemological foundation of segmentation analysis aligns directly with population thinking. Rather than seeking a single ``average'' profile, segmentation emphasizes carefully selecting segmentation identifiers to incorporate socialization pathways, including institutional positions, cultural resources, and social networks, thereby revealing the deep logic of social differentiation. This approach is particularly well-suited to LLM-based social simulation, that is, for LLMs to authentically simulate differential responses across groups in various social contexts, they must understand and reflect how social identity and positionality influences human viewpoints, reactions, and behaviors \parencite{Wang2025MisportrayFlatten}.

Segmentation is well suited to uncover heterogeneity that is masked by averaging. A compelling illustration comes from \textcite{Guenther2018PromisesReservations}'s study of South African public attitudes toward science and technology. Through a hierarchical cluster analysis on 3,183 respondents, they identified six subgroups characterized by distinct configurations of socioeconomic status, cultural capital, information sources, and science attitudes. These patterns were not visible in the overall averages. Only through segmentation analysis could the unique cultural distances and differentiated responses to science communication be revealed. This example highlights segmentation's ability to restore heterogeneity at an operational level—precisely what is needed to counteract the masking effects.

\subsection{Segmentation identifiers in prior LLM-based simulation research}
Although segmentation has not been systematically formalized as a theoretical concept in LLM-based social simulation, multiple studies have exhibited practical awareness of segmentation-style thinking. A review of existing work suggests that identifier selection roughly follows three logics borrowed from traditional segmentation research, though often without explicit theoretical justification.

First, demographic identifiers (e.g., age, gender, education, etc.) are the most commonly used segmentation identifiers, largely because they are easy to obtain and widely available in survey data \parencite{Slater1996TheoryMethod,Dibb2009ImplementationRules}. These variables dominate political science and public opinion simulation studies \parencite{Bisbee2024SyntheticReplacements,Suh2025FineTuningSurvey,Sun2024RandomSilicon}. For instance, \textcite{Argyle2023OutOfOneMany} combined GPT-3 with demographic background variables from the American National Election Studies through silicon sampling and found that model-generated ``silicon samples'' closely matched survey data on vote choice, party identification, and ideological distribution. This demonstrates that demographic characteristics can enhance simulation performance.

Second, psychographic identifiers (e.g., attitudes, values, and issue-specific cognitions) have been incorporated to address the limitation that demographics alone are often insufficient to distinguish attitudinal differences within seemingly homogeneous groups. For example, adding environmental policy stances and scientific-consensus cognitions to prompts improved predictive accuracy in climate change research \parencite{Lee2024EstimatePublicOpinion,Hine2014AudienceSegmentation}. Similarly, \textcite{Luo2024CulturalSensitivity} introduced the Hofstede individualism index as a cultural segmentation variable and showed that LLMs can capture not only demographic and psychological dimensions but also deep self-concept differences in cross-national settings.

Third, behavioral identifiers can be used to segment populations based on observed behavioral patterns. In human-computer interaction, \textcite{Lu2025AgentsActLikeUs} used user click, input, and submission paths to construct behavioral segments, enabling LLMs to reproduce interactive scenarios in ways that more closely reflect differences amongst users.

\subsection{From identifiers to research design: Key methodological challenges and considerations}
While these categories---from broad demographics to specific psychographic and behavioral traces---provide raw materials for persona conditioning, simply identifying them does not resolve the deeper implementation problem. Moving from these markers to an operational research design raises significant yet often overlooked methodological challenges.

The first challenge is identifier granularity. Existing LLM-based simulations tend to operationalize identifier inclusion as an all-or-nothing decision: Either a variable is incorporated into the persona prompt or it is omitted entirely \parencite{Argyle2023OutOfOneMany,Sun2024RandomSilicon,Lee2024EstimatePublicOpinion}. What these practices leave unaddressed is how much descriptive detail is needed to characterize a synthetic persona in ways that reflect true social positionality. This oversight stems from a failure to recognize that the restoration of heterogeneity is not merely a matter of whether to condition a model, but of how deeply to describe the social self. If we accept that social identity is rooted in layered social structures, simulation practices should move beyond ``thin'' descriptions based on isolated (demographic) labels toward ``thick'' descriptions that incorporate multidimensional identities. By conceptualizing descriptive detail as a spectrum of granularity, we can examine the point at which an LLM shifts from generating homogenized responses to capturing authentic subgroup variance. Accordingly, we ask the following research question (RQ):

\textbf{RQ1:} How does the granularity of segmentation identifiers—ranging from single---dimensional to layered, multidimensional conditioning---affect the performance of LLM-based simulation?

The second challenge, closely related to granularity, is the issue of quantitative parsimony. Shifting from ``adding identity information'' to ``designing segmentation as an experimental factor'' foregrounds a classic methodological trade-off: determining whether additional identifiers contribute meaningful signal or instead introduce redundancy, noise, or over-constraint instead. Emerging evidence \parencite{Anthis2025Position} indicates a non-linear relationship between informational complexity and simulation performance; while insufficient descriptive detail may produce ``average persona'' bias, indiscriminately accumulating segmentation identifiers risks introducing noise that obscures subgroup variance \parencite{Anthis2025Position}. Effective simulation therefore requires a balance---an ``informative parsimony''---that anchors the model without being over-regularized by irrelevant identifiers. Therefore, we ask the following research question:

\textbf{RQ2:} What trade-offs emerge between parsimony and comprehensiveness as the number of identifiers used for persona conditioning increases, and how do these manifest in simulation performance across evaluation metrics?

The third challenge concerns identifier selection logic. The field of LLM social simulation still lacks a unified framework for defining primary segmentation dimensions, making the rationale for choosing specific identifiers unclear and prone to inconsistent interpretation. For example, \textcite{Lee2024EstimatePublicOpinion} concatenated demographic characteristics with climate-related attitudinal variables to condition LLM predictions of public opinion. Although this improved predictive accuracy relative to demographics-only conditioning, such ad hoc data-driven configurations provide limited theoretical justification for why some identifiers are included while others are excluded. Without an explicit selection principle, performance gains may reflect opportunistic prompt tailoring that is practically useful but difficult to interpret as evidence of specific psychological or behavioral mechanisms. To move beyond case-by-case trial and error, identifier-selection logic should be made explicit and evaluated as an experimental factor.

In this study, we operationalized selection logic into three distinct approaches: theory-driven selection, purely data-driven selection, and an important middle path, pre-validated instrument-based selection. Under this intermediate logic, identifiers were drawn from established survey instruments that were developed through theory-informed design and subsequently validated on large samples. A representative example is the Yale Program on Climate Change Communication (YPCCC) ``Six Americas'' framework \parencite{Chryst2018SASSY}, which offers a validated segmentation instrument (question set) for assigning respondents to audience segments. By design, this approach is neither purely theoretical nor purely data-mined; it operationalizes theoretically meaningful constructs while retaining demonstrated predictive and classification performance. Therefore, we ask the following research question:

\textbf{RQ3:} How do theory-driven, prevalidated instrument–based, and data-driven identifier selection logics compare in their effects on simulation performance?

\section{Evaluating Heterogeneity Restoration}

To evaluate whether segmentation strategies restore population heterogeneity in theory-informed LLM-based social simulation, we use a comprehensive framework that diagnoses simulation performance across multiple dimensions. Existing studies have employed various metrics, but these methods were ad hoc and fragmented, with limited explicit connection to heterogeneity masking. Building on this gap, we propose a three-dimensional evaluation framework, informed by recent empirical studies \parencite{Argyle2023OutOfOneMany,Boelaert2025MachineBias,Bisbee2024SyntheticReplacements,Lee2024EstimatePublicOpinion,Luo2024CulturalSensitivity}, to operationalize heterogeneity restoration across distributional, structural, and predictive fidelity (see Table~\ref{tab:heterogeneity_framework}). Each dimension targets a distinct facet of the extent to which social variation is preserved or masked.

\begin{table}[htbp]
\centering
\caption{Three-dimensional framework for evaluating heterogeneity restoration}
\label{tab:heterogeneity_framework}
\begin{tabularx}{\textwidth}{@{}>{\raggedright\arraybackslash}p{0.22\textwidth}>{\raggedright\arraybackslash}p{0.30\textwidth}>{\raggedright\arraybackslash}X@{}}
\toprule
\textbf{Dimension} & \textbf{Theoretical focus} & \textbf{Key metrics} \\
\midrule
Distributional fidelity & Overall distribution alignment & Accuracy, Precision, Recall, F1 score; Mean Absolute Error (MAE); Kullback-Leibler divergence (KLD) \\
\midrule
Structural fidelity & Subgroup variance preservation & Within-group: Standard Deviation (SD), Coefficient of Variation (CV); Between-group: Normalized Earth mover's distance (nEMD), Multidimensional scaling (MDS), Procrustes distance \\
\midrule
Predictive fidelity & Relational pattern correspondence & Cramér's V \\
\bottomrule
\end{tabularx}
\end{table}

\subsection{Distributional fidelity}
Distributional fidelity examines whether LLM-generated responses match the empirical distributions of the outcome variables in terms of both central tendency and distributional shape.Many prior studies have relied on mean-based comparisons to evaluate the aggregate similarity between simulated and human responses. For ordered categorical outcomes, especially Likert-style responses, the mean absolute error (MAE) is a useful complementary measure because it captures the absolute numerical deviation between simulated and empirical responses when response categories are treated as quasi-continuous. When responses are evaluated as discrete categories, classification metrics such as accuracy, precision, recall, and F1 score are used to assess agreement between simulated and empirical responses across response options. 

It is noteworthy, however, that agreement in means or category-level allocations does not necessarily indicate recovery of the full distributional structure. Prior work similarly cautions that matching aggregate statistics can still miss subgroup-level distributional patterns \parencite{Boelaert2025MachineBias}. This suggests that LLMs may fit aggregate statistics by generating ``average'' individuals, completely missing meaningful subgroup clusters \parencite{Wu2025Boundary}. To address this limitation, distribution distance metrics such as Kullback-Leibler divergence (KLD), which quantifies the information loss when one distribution approximates another, offer a complementary assessment.

\subsection{Structural fidelity}
Structural fidelity evaluates whether differences in variance across social subgroups, together with the relational patterns within and between them, are accurately preserved. This dimension directly addresses heterogeneity masking, specially, the tendency of LLMs to homogenize responses across social subgroups that should exhibit distinct patterns of variation.

Structural fidelity can be decomposed into two complementary aspects (within-group variance and between-group structure). Within-group variance examines whether subgroups maintain internal diversity. This aspect is assessed using subgroup-level standard deviation (SD) and the coefficient of variation (CV = SD/Mean), which capture the degree of internal variation within social segments. These values are then compared with human benchmarks to evaluate whether LLM-generated subgroups are overly homogeneous.

Between-group structure evaluates whether LLMs preserve systematic differences across potential social segments. We use the normalized Earth mover's distance (nEMD) to quantify discrepancies between distributions across all subgroup pairs; higher values indicate stronger sensitivity of model outputs to subgroup labels \parencite{Boelaert2025MachineBias}. Additionally, multidimensional scaling (MDS) is used to represent subgroup relations in a two-dimensional relational (identity) space, while Procrustes distance provides a complementary measure of alignment between simulated and empirical subgroup structures.

\subsection{Predictive fidelity}
Social science is fundamentally concerned with understanding relationships between variables and phenomena rather than merely describing them in isolation. Predictive fidelity extends beyond simple distributional fit to examine relational pattern correspondence, that is, whether LLMs can accurately capture and reproduce the relational logic linking social positions (x) to attitudinal outcomes (y) as observed in real-world data. This dimension evaluates the extent to which LLMs have internalized authentic associational patterns that enable meaningful differentiation.

Widely used measures of relational structure encompass correlation coefficients, such as Pearson's correlation for continuous variables and Spearman's correlation for ordinal data, as well as tetrachoric correlation for binary variables \parencite{Argyle2023OutOfOneMany}. Cramér's V serves as a primary diagnostic for categorical associations that quantifies whether models faithfully reproduce structured preference patterns. By comparing Cramér's V values for the same variable pairs (e.g., political affiliation and climate attitudes) across human and simulated responses, researchers can assess whether models preserve authentic relational structures or collapse responses into homogenized averages. If human data show moderate association but simulations produce near-deterministic patterns, this signals over-regularization, that is, the model has imposed exaggerated dependencies unsupported by empirical reality. Ultimately, evaluating predictive fidelity reveals whether models learn genuine relational patterns or merely spurious associations. However, high relational correspondence indicates successful ``heterogeneity restoration'' only when both distributional and structural fidelity are adequately established.

In summary, this three-dimensional framework integrates commonly used metrics from the LLM simulation literature into a coherent structure specifically designed to diagnose heterogeneity restoration. The dimensions are mutually reinforcing: Distributional fidelity captures whether simulated outputs reproduce aggregate response patterns, structural fidelity assesses whether subgroup heterogeneity and between-group structure are preserved, and predictive fidelity evaluates whether these subgroup differences support meaningful downstream associations.

In what follows, we employ this evaluation framework to systematically assess the simulation performance of various segmentation configurations along three key dimensions: granularity (the richness of information included), parsimony (the number of segmentation identifiers), and variable selection logic (theory-driven, data-driven, or pre-validated instrument-based approaches). 

\section{Methodology}
\subsection{Data}
We employ a comparative design centered on climate opinion attitudes in the United States, leveraging both human and synthetic “silicon sample” data. Human benchmark samples were collected in October 2025 through carefully stratified surveys distributed via Prolific, a leading online research platform trusted for diverse and representative participant pools. Sampling quotas were aligned with national census benchmarks to ensure demographic representativeness across key dimensions, including gender, age, and region. Strict protocols were implemented: Only unique (via IP filtering), attentive respondents were included, with invalid submissions filtered via response-time metrics and manual inspection. Following quality control, we retained 594 valid samples for further analysis. Respondents’ baseline segment memberships were assigned using the Yale Program on Climate Change Communication's Six Americas Super Short Survey, which classifies individuals into one of six audience segments based on four climate-related attitudinal items. The six segments are alarmed, concerned, cautious, disengaged, doubtful, and dismissive (for details, see \parencite{Chryst2018SASSY}). In our survey, each respondent completed the four-item questionnaire, which we used to assign an individual-level segment label.

For synthetic silicon samples, we employed Llama 3.1-70B and Mixtral 8x22B, which were selected as high-performing open-weight models available at the time of data collection after benchmarking several alternatives (e.g., Qwen, Gemma3, and Bloom) against simulation-relevant criteria. Our criteria included (1) task completion success, (2) the absence of extraneous assistant commentary, (3) internal logical consistency under persona conditioning, and (4) diversity of attitudinal expression \parencite{Lyman2025BalancingAlignment}. These criteria, especially criterion 4, were deliberately designed to prioritize heterogeneity preservation rather than merely average performance. LLM completions were then systematically generated using persona conditioning aligned with each segmentation configuration.

\subsection{Segmentation configurations} \label{subsec:Segmentation_configurations}
To systematically examine how segmentation choices affect simulation performance, we manipulated three aspects of segmentation identifiers. Detailed overviews of the six configurations and their exact identifier wording are provided in Appendix A2–A3 (Tables A2--A8):

\textbf{Granularity.} This aspect captures the richness of information used to construct persona prompts. We operationalized granularity by identifying a core set of identifiers from different dimensions (sociodemographic, psychological, and behavioral), and then progressively increasing the amount and dimensional breadth of conditioning information. These combinations were then used to craft prompts at varying levels of granularity, enabling us to parse the incremental simulation performance gains from including finer-grained, multidimensional segmentation identifiers. This design isolates how layered social identity information translates into more nuanced LLM subgroup distinctions and their contributions to improve or undermine simulation performance.

\textbf{Parsimony.} This aspect addresses the trade-off between comprehensiveness and parsimony in selecting segmentation identifiers. We systematically compared the impact of multi-item identifiers with that of more streamlined, parsimonious identifiers. The comprehensive condition included a wide array of identifiers from each dimension (i.e., sociodemographic, psychographic, and behavioral), whereas the parsimonious condition was distilled into a reduced core subset of high-impact identifiers. We tested whether additional identifiers yielded diminishing or negative returns in terms of simulation performance.

\textbf{Identifier selection logic.} This aspect contrasts with three distinct methodological approaches to selecting which identifiers to include in segmentation profiles. \textit{Theory-driven} condition draws upon established theoretical models of climate attitudes and behavioral change, selecting identifiers with direct theoretical justification and prior empirical validation. In contrast, \textit{data-driven} condition prioritizes statistical efficiency, applying machine learning techniques (e.g., Gradient Boosting Machines, GBM) to rank candidate identifiers and select those with the greatest discriminative power for segment differentiation. Given that our data focus on attitudes toward climate change, we further incorporated an empirical logic (i.e., a \textit{prevalidated instrument–based} condition) that leverages the standardized SASSY instrument \parencite{Chryst2018SASSY}. There are two versions of SASSY for audience segmentation, i.e., a detailed 15-item instrument (Item-15) and a shorter four-item version (Item-4). We implemented both versions for comparison. The comparison across and within identifier selection approaches aimed to elucidate the trade-offs between theoretical interpretability, statistical performance, and efficiency in constructing effective segmentation strategies.

Based on these three aspects, we derived six segmentation configurations. We addressed the three RQs by juxtaposing and comparing these configurations (see Table~\ref{tab:segmentation-configurations} for details). The configurations are as follows:

\begin{itemize}
  \item \textit{Demo:} Baseline profiles built exclusively from standard demographic variables (i.e., gender, age, education, income, and ethnicity), paralleling common social simulation defaults.
  \item \textit{Demo+Theory-59:} Profiles integrating demographic variables with 59 theoretically derived psychological and behavioral indicators relevant to climate attitudes, providing the richest and most comprehensive configuration.
  \item \textit{Demo+Theory-15:} A streamlined, reduced version of the theory-informed profile including only the 15 most consequential predictors as identified by prior climate communication research.
  \item \textit{Data-driven:} A profile using the top 15 high-impact identifiers as determined by gradient boosting machine-based feature selection, reflecting the data-driven approach to optimal subgroup discrimination.
  \item \textit{Item-15:} Replicates the SASSY instrument, using 15 empirically validated climate-related attitudinal questions/items as identifiers (i.e., climate change beliefs, risk perceptions, and policy preferences) for segmentation.
  \item \textit{Item-4:} Applies a condensed four-item version of the SASSY instrument for efficient, low-burden segmentation.
\end{itemize}

\begin{table}[htbp]
\centering
\caption{Segmentation configurations}
\label{tab:segmentation-configurations}
\footnotesize
\setlength{\tabcolsep}{4pt}
\begin{tabularx}{\textwidth}{@{}>{\raggedright\arraybackslash}p{0.19\textwidth}>{\raggedright\arraybackslash}p{0.13\textwidth}>{\raggedright\arraybackslash}p{0.22\textwidth}>{\raggedright\arraybackslash}X>{\raggedright\arraybackslash}p{0.12\textwidth}@{}}
\toprule
\textbf{Segmentation configuration} & \textbf{Granularity} & \textbf{Parsimony} & \textbf{Identifier selection logic} & \textbf{Addressed RQs}\\
\midrule
Demo & Low (single-dimensional) & High (5 identifiers) & Empirical default & Baseline for comparison \\
Demo+Theory-59 & Very high (multidimensional) & Low (59 identifiers) & Theory-driven & RQ1, RQ2 \\
Demo+Theory-15 & High (multidimensional) & Medium (15 identifiers) & Theory-driven & RQ1--RQ3 \\
Data-driven & High (multidimensional) & Medium (15 identifiers) & Data-driven (GBM) & RQ3 \\
Item-15 & Medium (attitude-focused) & Medium (15 identifiers) & Pre-validated instrument-based & RQ2, RQ3 \\
Item-4 & Medium (attitude-focused) & High (4 identifiers) & Pre-validated instrument-based & RQ2, RQ3 \\
\bottomrule
\end{tabularx}
\end{table}

\subsection{Prompting strategies and responses} \label{subsec:Prompting_strategies}

All simulations were conducted using a zero-shot, Q\&A prompt template. This format was chosen to construct an immersive context for the model to adopt a specified persona, while the zero-shot approach prevents the introduction of confounding variables that can arise from few-shot exemplars \parencite{Santurkar2023WhoseOpinions}. To encourage response diversity while maintaining coherence, a decoding temperature of 0.8 was employed in conjunction with a top-p value of 1.0 to prevent artificially constraining the output distribution \parencite{Argyle2023OutOfOneMany}.

Each prompt elicited attitudinal responses (i.e., outcome items in this study) toward climate change under specified persona conditions. Specifically, the model was asked to rate its perception of climate change on three seven-point Likert scales: (1) the degree to which climate change is perceived as pleasant or unpleasant (\textit{pleasant}), (2) its overall favorability or unfavourability (\textit{favorable}), and (3) its general positivity or negativity (\textit{positivity}). The corresponding questionnaire items are Q25, Q26, and Q27, respectively (see Table A1 in Appendix A for details). The model’s task was to respond to each item by providing only the corresponding integer value (1–7), thereby ensuring standardized quantitative outputs suitable for statistical comparison with human survey data.

\subsection{Evaluation metrics}
To evaluate how different segmentation strategies restore heterogeneity, we operationalized our evaluation framework along three complementary dimensions—distributional, structural, and predictive fidelity—and applied it consistently to all six segmentation configurations.

Distributional fidelity was assessed using MAE, accuracy, precision, recall, F1 score, and KLD. MAE captured outcome item–level numerical deviations between empirical and simulated mean responses, treating the 7-point Likert scale as quasi-continuous. Accuracy, precision, recall, and F1 score summarized agreement patterns at the response-category level (treating each Likert point as a discrete class), whereas KLD quantified information loss when the simulated distribution approximated the empirical response distribution. Because outcome variables were measured on a 7-point Likert scale and empirical response categories were highly imbalanced, we report weighted precision, weighted recall, and weighted F1 score to ensure that performance reflected observed category frequencies rather than being driven by rare response options. For each outcome item, we computed these measures by comparing the empirical frequency distribution with the corresponding LLM-generated frequency distribution and then summarized the results across outcomes to obtain configuration-level estimates of distributional fidelity.

Structural fidelity was assessed at two levels: within-group variance and between-group structure. For within-group variance, we calculated subgroup-level SD and CV and compared these values with the corresponding empirical human benchmarks. Between-group structure was evaluated separately for each segmentation configuration and each LLM. Within each configuration, we represented each subgroup with its outcome distribution and computed nEMD for each subgroup pair, yielding a subgroup-by-subgroup distance matrix for the empirical data and a corresponding distance matrix for each simulated LLM output. Pairwise nEMD captured distributional differences between subgroups. To summarize between-group structure, we first computed the median pairwise nEMD within each outcome item and then averaged these outcome item–level medians across the outcome variables to obtain a configuration-level estimate. To assess whether subgroup differences were organized similarly in the empirical and simulated data, we applied MDS to each distance matrix. MDS projects each subgroup distance matrix into a two-dimensional map in which subgroups with response distributions that are more similar appear closer together, whereas subgroups that are more distinct appear farther apart. In this way, the MDS map provides a visual representation of the overall pattern of subgroup relationships. Because two MDS maps could represent the same underlying structure while differing in orientation, position, or scale, they could not be compared directly using raw coordinates. We therefore used a Procrustes transformation to align each simulated map with the empirical map before comparison. The resulting Procrustes distance quantified the remaining mismatch between the two subgroup structures after these geometric differences were removed.

Predictive fidelity examined whether segmentation identifiers retained their empirical association with outcome items in the simulated data. We operationalized this dimension via Cramér’s V, computed from contingency tables between each segmentation identifier and each outcome item. Cramér’s V provides a standardized effect-size estimate of association strength, enabling direct comparison of whether a configuration underestimates, matches, or inflates empirical subgroup–outcome linkages. We summarize predictive fidelity by aggregating association estimates across identifiers and outcomes.

By jointly evaluating segmentation strategies across distributional, structural, and predictive fidelity, we were able to distinguish configurations that yielded comprehensive heterogeneity restoration from those that merely optimized a single metric while degrading other properties of the empirical system. 

\section{Results}
This section evaluates simulation performance along three complementary dimensions. We present the main findings in relation to the three RQs.

\subsection{RQ1: Increasing granularity does not produce consistent improvements in simulation performance}

To answer RQ1, we compared three segmentation configurations: Demo, Demo+Theory-15, and Demo+Theory-59.

In terms of distributional fidelity (see Appendix A4, Table 9 and Figure~\ref{fig:distributional-fidelity}), the results show that increasing granularity does not result in uniform improvement across metrics. Moving from Demo to Demo+Theory-15 generally improved performance, especially in KLD and several category-level classification metrics. However, moving further to Demo+Theory-59 did not provide additional gains and, in some respects, worsened performance.

The Demo configuration performed worst in distributional shape for both LLMs (KLD: 2.72 for Llama and 6.58 for Mixtral) and yielded the lowest weighted F1 scores (.46 and .39, respectively), indicating that demographics-only conditioning departs substantially from the empirical distribution. Demo+Theory-15 improved this baseline in both LLMs. For Llama, KLD decreased from 2.72 to .68, F1 increased from .46 to .51, and accuracy/recall also improved, although MAE rose modestly from .38 to .46. For Mixtral, MAE decreased from .89 to .80, KLD decreased from 6.58 to 1.51, and F1 increased from .39 to .40.

Cross-LLM averages reinforced this pattern. Relative to Demo, Demo+Theory-15 reduced average KLD from 4.65 to 1.10 and increased average F1 from .43 to .46, while average MAE remained nearly unchanged (.64 vs. .63). In contrast, Demo+Theory-59 showed weaker average performance (KLD = 2.76, F1 = .45, MAE = .69). Category-level metrics changed only modestly across configurations: Average accuracy increased from .52 (Demo) to .555 (Demo+Theory-59) and .560 (Demo+Theory-15); average precision from .39 to .40 and .42; average recall from .52 to .555 and .56; and average F1 from .425 to .445 and .455. Taken together, these comparisons indicate that additional granularity beyond the intermediate level did not improve overall distributional fidelity.

\begin{figure}[htbp]
\centering
\includegraphics[width=\textwidth]{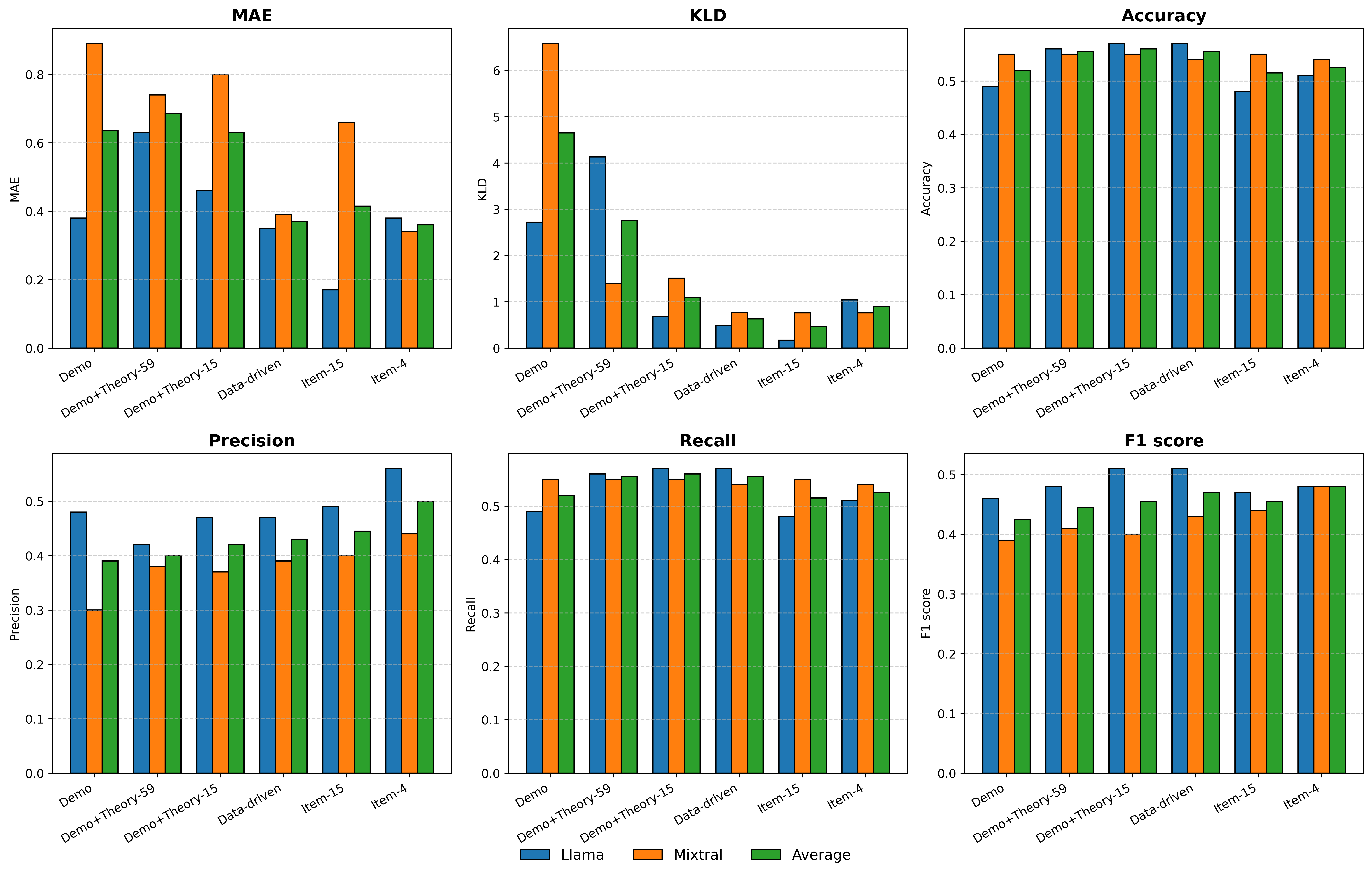}
\caption{Comparison of distributional fidelity metrics across segmentation configurations and LLMs.}
\label{fig:distributional-fidelity}
\end{figure}

In terms of structural fidelity (see Appendix A4, Table 10 and Figure~\ref{fig:structural-fidelity}), the results suggest that intermediate granularity (Demo+Theory-15) generally outperformed both the Demo baseline and the more granular Demo+Theory-59 configuration. Specifically, Demo+Theory-15 preserved substantially more within-group variance than the Demo configuration in both LLMs: SD increased from .46 (Demo) to .69 (Demo+Theory-15) in Llama and from .02 to .60 in Mixtral, with CV rising correspondingly from .27 to .37 and from .02 to .39. On average across both LLMs, SD increased from .24 under the Demo configuration to .63 under Demo+Theory-15, and CV rose from .14 to .38. In contrast, the Demo+Theory-59 configuration did not sustain this improvement. In Llama, SD fell back to .44 and CV to .34, both below the Demo+Theory-15 configuration and closer to the Demo baseline; in Mixtral, SD and CV shifted to .63 and .32, respectively. Cross-LLM averages also revealed that Demo+Theory-15 slightly outperformed the Demo+Theory-59 configuration (SD: .54; CV: .33), suggesting that adding more identifiers did not consistently improve the recovery of within-group variance.

\begin{figure}[htbp]
\centering
\includegraphics[width=\textwidth]{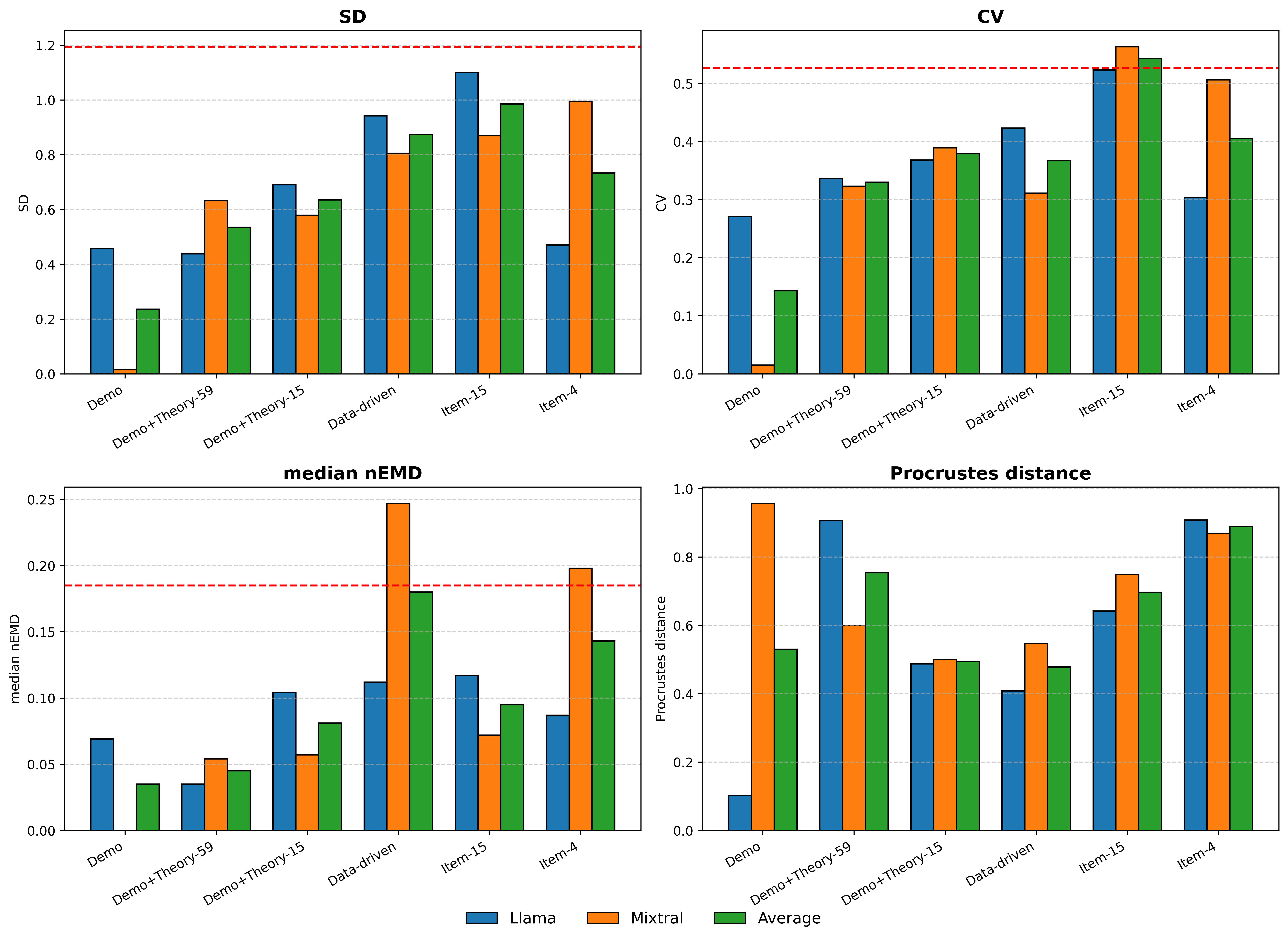}
\caption{Comparison of structural fidelity metrics across segmentation configurations and LLMs.}
\label{fig:structural-fidelity}
\end{figure}

\begin{figure}[htbp]
\centering
\includegraphics[width=\textwidth,height=0.82\textheight,keepaspectratio]{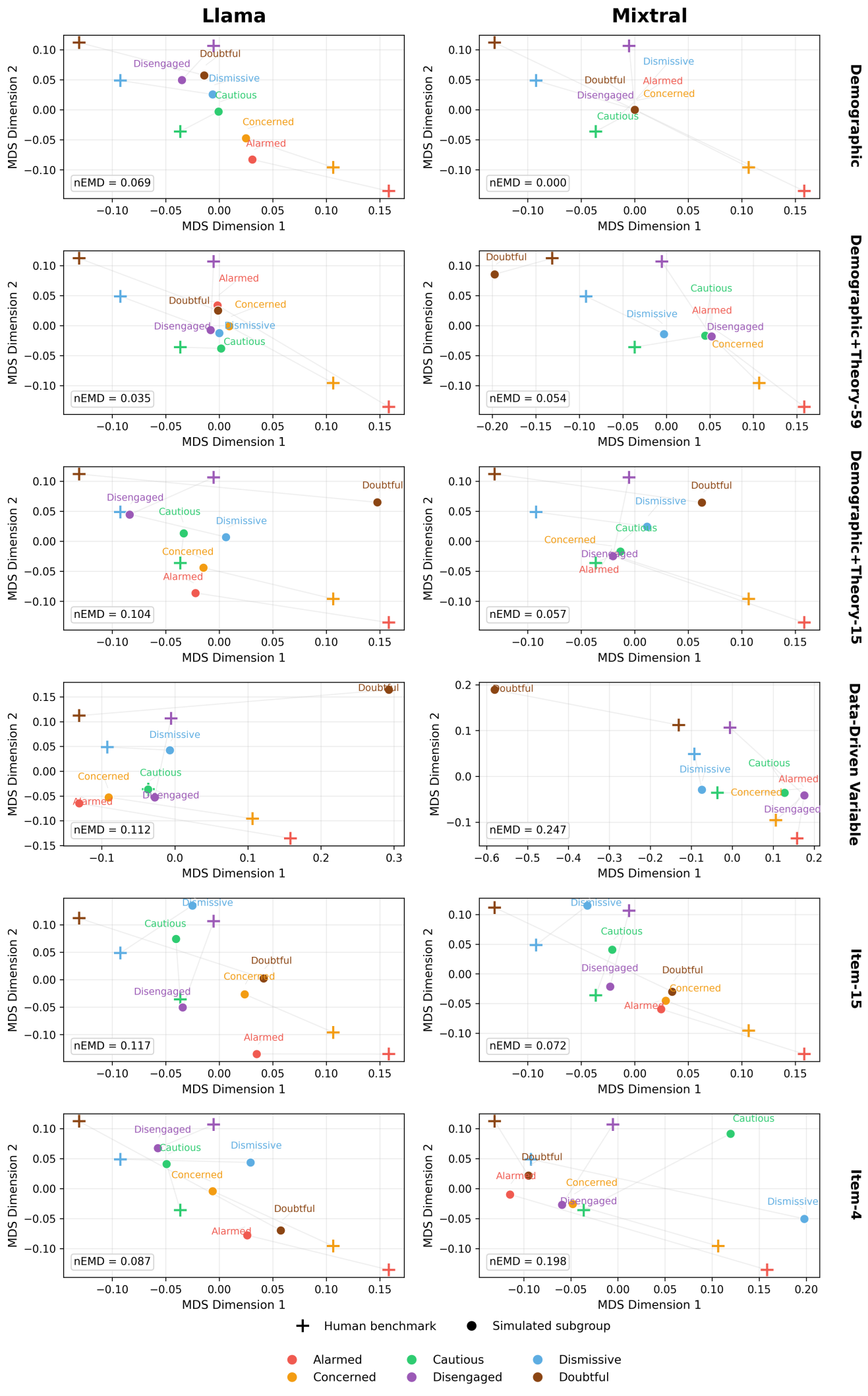}
\caption{MDS maps of empirical and simulated subgroup structures across segmentation configurations and LLMs.}
\begin{flushleft}
\footnotesize \textit{Note.} Colors denote subgroup identity and are held constant across panels, allowing direct comparison between empirical and simulated subgroup locations. Distances between points reflect differences in subgroup response distributions derived from pairwise nEMD, such that closer points indicate more similar subgroup distributions. The reported nEMD in each panel summarizes the overall magnitude of between-group structure within that segmentation configuration. MDS Dimensions 1 and MDS Dimensions 2 do not have direct substantive interpretations; thin gray lines connect each simulated subgroup to its empirical counterpart only to facilitate visual comparison and should not be interpreted as between-group distances themselves.
\end{flushleft}
\label{fig:mds}
\end{figure}

Between-group structure exhibited a similar limit to increasing granularity. Although the Demo+Theory-15 configuration (median nEMD = .10 for Llama, .06 for Mixtral; average = .08) and the Demo+Theory-59 configuration (median nEMD = .04 for Llama, .05 for Mixtral; average = .04) yielded relatively higher median nEMD values than the Demo configuration (median nEMD = .07 for Llama, .00 for Mixtral; average = .03), both remained well below the human benchmark of .19. This pattern indicated that subgroup pairs in the simulations remained less differentiated from one another than they were empirically, and that increasing granularity beyond an intermediate level did not strengthen between-group structure. Figure~\ref{fig:mds} complemented Appendix A4, Table 10 by visualizing these subgroup patterns directly against the human benchmark. Procrustes distance provided an additional summary of how closely the simulated subgroup map resembled the empirical one. In Llama, Procrustes distance rose from .10 under the Demo baseline to .49 under Demo+Theory-15 and to .91 under Demo+Theory-59; in Mixtral, it decreased from .96 to .50 and then increased again to .60. While the LLM-specific baseline distances varied, the cross-LLM average Procrustes distance improved from .53 under the Demo baseline to .49 under Demo+Theory-15; however, expanding to Demo+Theory-59 worsened the average distance to .75. Taken together, these results suggested that increasing the granularity of segmentation identifiers beyond an informative threshold did not bring the simulated subgroup structure closer to the empirical one; instead, subgroup differences remained weaker than in the human data, and the overall subgroup pattern became less stable.

\begin{figure}[htbp]
\centering
\includegraphics[width=\textwidth]{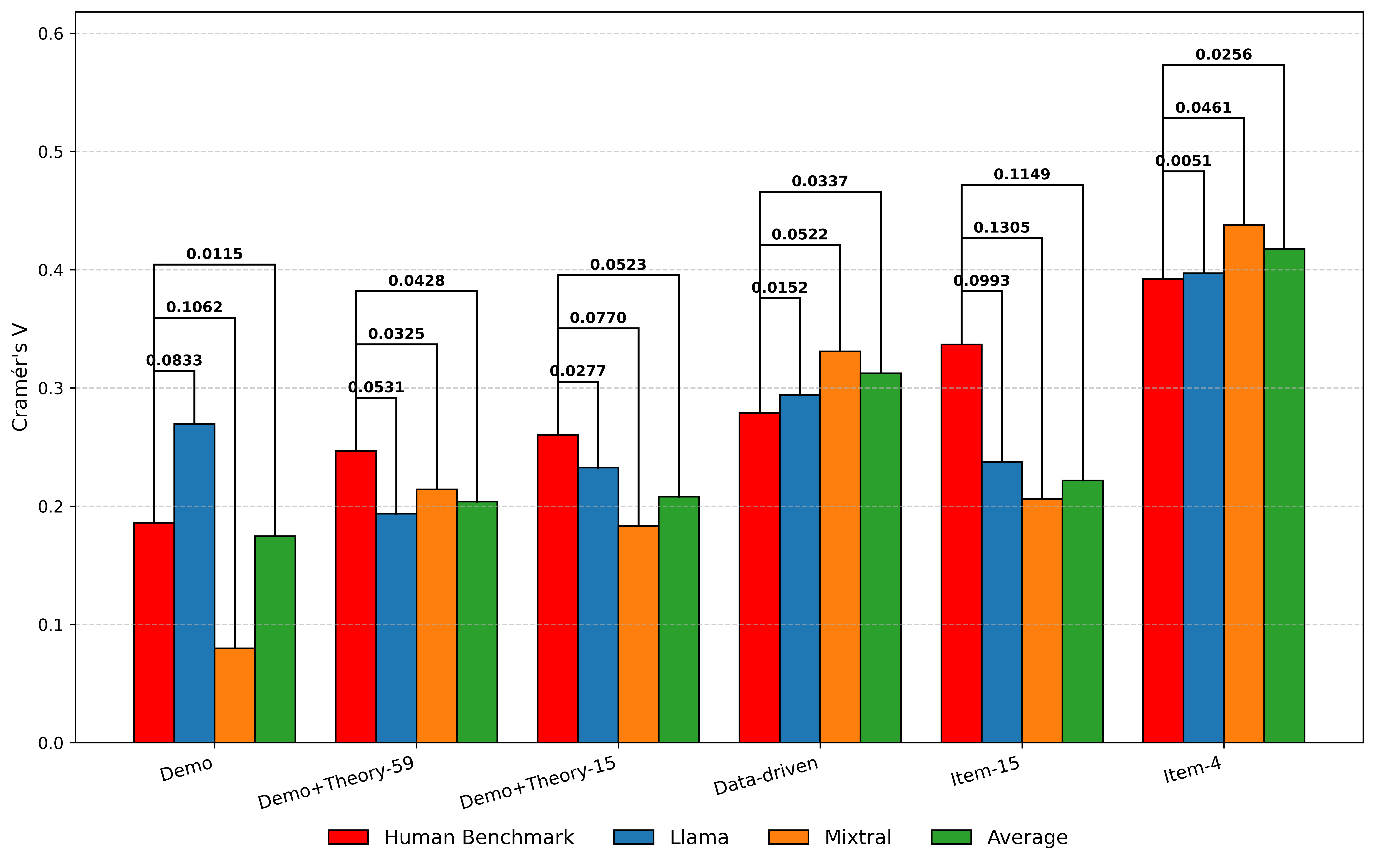}
\caption{Comparison of Cramér's V across segmentation configurations and LLMs. Brackets and numbers indicate the differences from the human benchmark.}
\label{fig:cramersv}
\end{figure}

In terms of predictive fidelity, Table~\ref{tab:cramers_v} and Figure~\ref{fig:cramersv} show that, among the demographic and theory-based configurations, Demo was consistently furthest from its human benchmark in both LLMs, with a cross-LLM average simulated Cramér's V of .17 against a human benchmark that ranged from .19 to .26. The relative performance of the Demo+Theory-15 and Demo+Theory-59 configurations, however, was unstable across LLMs. In Llama, Demo+Theory-15 produced a simulated Cramér's V of .23 against a human benchmark of .26 (a difference of .03), while Demo+Theory-59 reached only .19 against its benchmark of .25 (a difference of .05), making the intermediate configuration the better one. In Mixtral, this order is reversed: Demo+Theory-59 yielded a smaller difference from the human benchmark (.03) than Demo+Theory-15 (.08). Overall, these results indicated that the point at which additional identifiers contributed informative signals rather than noise was not fixed across LLMs and could not be naively determined by the granularity of identifiers.

\begin{table}[htbp]
\centering
\caption{Summary of Cramér's V values}
\label{tab:cramers_v}
\begin{tabularx}{\textwidth}{@{}>{\raggedright\arraybackslash}Xrrr@{}}
\toprule
\textbf{Segmentation configuration} & \textbf{Human benchmark} & \textbf{Llama} & \textbf{Mixtral} \\
\midrule
Demo                & .19 & .27 & .08 \\
Demo+Theory-59      & .25 & .19 & .21 \\
Demo+Theory-15      & .26 & .23 & .18 \\
Data-driven         & .28 & .29 & .33 \\
Item-15             & .34 & .24 & .21 \\
Item-4              & .39 & .40 & .44 \\
\bottomrule
\end{tabularx}
\end{table}

% This cross-LLM instability is also visible when the two LLMs are averaged. The cross-LLM aggregated Cramér's V is .21 for Demo+Theory-15 and .20 for Demo+Theory-59, but since these configurations have different human benchmarks, the more informative comparison is their average absolute deviation from those benchmarks: .05 for Demo+Theory-15 versus .04 for Demo+Theory-59. By contrast, Demo configuration remains clearly least accurate, with a larger average absolute deviation of 0.0948 across the two LLMs. 

% Furthermore, Figure~\ref{fig:item_cramersv_theory59} suggests that the apparent advantage of Demo+Theory-59 in Mixtral does not hold consistently at the outcome item-level, where individual identifier-outcome associations remain poorly aligned with their empirical counterparts despite the closer aggregate mean. O

% \begin{figure}[htbp]
% \centering
% \includegraphics[width=\textwidth,height=0.82\textheight,keepaspectratio]{item_cramers_v_Theory59.png}
% \caption{Cramér's V under Demo+Theory-59 segmentation configuration}
% \label{fig:item_cramersv_theory59}
% \end{figure}

\subsection{RQ2: Informative parsimony outperforms comprehensiveness}

To address RQ2, we compared two pairs of configurations: a theory-driven pair (Demo+Theory-15 vs. Demo+Theory-59) and a pre-validated instrument-based pair (Item-4 vs. Item-15). Because the theory-driven contrast has already been discussed in the previous section, here, we focus here on the instrument-based comparison

Regarding distributional fidelity, the instrument-based pair exhibited a more conditional pattern than the theory-driven pair (see Appendix A4, Table 9 and Figure~\ref{fig:distributional-fidelity}). Item-15 clearly improved KLD in Llama (.17 vs. 1.04 for Item-4) and matched Item-4 in Mixtral (both .76), yielding a lower cross-LLM average KLD overall (.47 for Item-15 vs. .90 for Item-4). However, MAE showed the opposite pattern across LLMs: Item-15 was better in Llama (.17 vs. .38), whereas Item-4 was better in Mixtral (.34 vs. .66), and Item-4 was slightly better in terms of the cross-LLM average (.36 vs. .42). Category-level classification metrics differed only modestly between the two configurations (all differences $\leq$ .08), indicating that the main distributional contrast was concentrated in KL divergence and MAE. Overall, these results suggested that parsimony did not inherently impair distributional fidelity compared to more comprehensive configurations.

Structural fidelity (see Appendix A4, Table 10 and Figure~\ref{fig:structural-fidelity}) did not mirror the pattern observed for distributional fidelity. In general, Item-15 better recovered within-group variance, whereas Item-4 tended to preserve somewhat stronger between-group differentiation; neither configuration reliably recovered between-group relational geometry. For within-group variance, Item-15 was consistently closer to the human benchmark (SD = 1.19, CV = .53) than Item-4 in both LLMs. The contrast was strongest in Llama, where Item-15 reached SD = 1.10 and CV = .52, compared with SD = .47 and CV = .30 for Item-4. Cross-LLM averages showed the same pattern: Item-15 reached SD = .99 and CV = .54, whereas Item-4 reached SD = .74 and CV = .41. For between-group structural differentiation, Item-4 was slightly stronger overall between-group structural differentiation, with a cross-LLM average median nEMD of .15, compared with .10 for Item-15. In Mixtral, Item-4 (.20) was also the only configuration of the two to reach the human benchmark (.19), whereas Item-15 remained far below (.07). In Llama, both remained below the benchmark (Item-15: .12; Item-4: .09). Procrustes results (see Figure~\ref{fig:mds}) further showed weak relational recovery between simulated and empirical subgroup structures.

Relational recovery between simulated and empirical subgroups remained weak: Distances remained high for both segmentation configurations (Item-15: .64–.75; Item-4: .87–.91), with cross-LLM averages of .70 and .89, respectively. Taken together with the theory-driven pair results, these findings indicated that greater identifier comprehensiveness did not yield consistent structural gains; more broadly, structural performance appeared to depend not simply on the number of added identifiers, but on whether those identifiers carried discriminative information relevant to subgroup structure.

Predictive fidelity showed the clearest advantage of Item-4 over Item-15 (see Table~\ref{tab:cramers_v} and Figure~\ref{fig:cramersv}). Under the Item-4 configuration, the human benchmark was .39, while the simulated Cramér's V values were .40 (Llama) and .44 (Mixtral), corresponding to gaps of .01 and .05, respectively. Under the Item-15 configuration, the human benchmark was .34, while the simulated Cramér's V values were .24 (Llama) and .21 (Mixtral), with much larger gaps of .10 and .13. At the aggregated level, the cross-LLM average gap was therefore substantially smaller for Item-4 (.03) than for Item-15 (.11), indicating better preservation of empirical identifier–outcome associations under the more parsimonious configuration. Combined with the theory-driven comparison—where Demo+Theory-59 also departed further from the benchmark than Demo+Theory-15, at least in Llama—these findings suggested that adding identifiers beyond an informative threshold might impair predictive fidelity rather than improve it.

In summary, all the evidence for RQ2 therefore lent stronger support to informative parsimony than to comprehensiveness; in other words, adding more identifiers did not reliably improve simulation performance, whereas a more compact and informative segmentation configuration tended to yield more robust overall performance.

\subsection{RQ3: The logic of identifier selection matters}

To answer RQ3, we compared three segmentation configurations that differed in identifier selection logic while holding the number of identifiers constant: Demo+Theory-15, Data-driven, and Item-15. Across the theory-driven, data-driven, and prevalidated instrument–based selection logics, the resulting configurations exhibited distinct fidelity profiles. It was found that in general, no single configuration performed best across all three evaluative dimensions and that the dimension that benefited most depended on how the identifiers were selected. 

In terms of distributional fidelity, no single selection logic dominated all metrics (see Appendix A4, Table 9 and Figure~\ref{fig:distributional-fidelity}). Item-15 performed best on KLD in both LLMs (.17 in Llama; .76 in Mixtral), yielding the lowest cross-LLM average KLD (.47), compared with Data-driven (.63) and Demo+Theory-15 (1.10). Moreover, the Item-15 and Data-driven configurations performed better on MAE overall, with a lower cross-LLM average MAE (.42 for Item-15; .37 for Data-driven) than Demo+Theory-15 (.63). Category-level classification metrics were relatively similar across the three configurations in both LLMs, with accuracy ranging from .48 to .57 and F1 score from .40 to .51. Taken together, these results suggested that the pre-validated instrument-based segmentation configuration provided the strongest distributional alignment with the empirical data.

Structural fidelity (see Appendix A4, Table 10 and Figure~\ref{fig:structural-fidelity}) did not indicate a single uniform winner across all metrics. Instead, the three configurations showed different strengths. For within-group variance, Item-15 performed best overall. Its cross-LLM averages (SD = .99, CV = .54) were closest to the human benchmarks (SD = 1.19, CV = .53), compared with Data-driven (SD = .87, CV = .37) and Demo+Theory-15 (SD = .64, CV = .38). Data-driven was intermediate in terms of SD but remained farther from the benchmark in terms of CV, especially in Mixtral (CV = .31). For between-group structure, the pattern reversed. Data-driven showed the strongest differentiation overall, with a cross-LLM average median nEMD of .18, which was closest to the human benchmark (.19), while Demo+Theory-15 and Item-15 remained lower (.08 and .10, respectively). This advantage was driven mainly by Mixtral, wherein Data-driven reached .25; in Llama, all three configurations remained below the benchmark (Demo+Theory-15: .10, Data-driven: .11, Item-15: .12). Geometric alignment (i.e., structural relation between simulated and empirical subgroups) showed a similar tendency: Data-driven had the lowest cross-LLM average Procrustes distance (.48), compared with Demo+Theory-15 (.49) and Item-15 (.70). Taken together, these results suggested that the Data-driven configuration yields the strongest between-group structural recovery, whereas Item-15 better captures within-group variance.

Predictive fidelity revealed a different pattern (see Table~\ref{tab:cramers_v} and Figure~\ref{fig:cramersv}). Because the three segmentation configurations had different human benchmark Cramér’s V values (Data-driven: .28; Demo+Theory-15: .26; Item-15: .34), the most informative comparison was each configuration’s absolute deviation from its own benchmark. Data-driven showed the smallest gaps in both LLMs (Llama: .02; Mixtral: .05), while Demo+Theory-15 had an intermediate gap (Llama: .03; Mixtral: .08), and Item-15 had the largest gap (Llama: .10; Mixtral: .13). The cross-LLM average gaps confirmed this pattern: .03 for Data-driven, .05 for Demo+Theory-15, and .12 for the Item-15 configuration. These results indicated that the Data-driven configuration best preserved empirical identifier–outcome association strength, whereas the Item-15 configuration failed to recover its higher benchmark associations despite having the highest empirical Cramér’s V.

Taken together, the evidence across the three fidelity dimensions indicated that identifier selection logic determined which dimension benefited most: Item-15 best preserved distributional shape, Data-driven best recovered between-group structure and identifier–outcome associations, and Demo+Theory-15 showed intermediate performance without a clear advantage in any single evaluative dimension.

\subsection{Summary of findings}

Across RQ1–RQ3, the results show that simulation performance depends less on the sheer number of identifiers than on whether those identifiers are informative in terms of the target task. First, the single-dimensional demographics-only configuration was an unreliable foundation across all three fidelity dimensions. It produced the greatest distributional divergence, weakened subgroup structure, and distorted association patterns. However, it is noteworthy that increasing granularity did not produce consistent improvements. Moving from Demo to Demo+Theory-15 improved several metrics, but further expansion to Demo+Theory-59 did not consistently improve performance and worsened structural and predictive fidelity. Second, the parsimony–comprehensiveness trade-off is fidelity specific. Parsimonious configurations often match or outperform more comprehensive alternatives, especially for structural and predictive fidelity, while distributional fidelity remains mixed depending on the metric (e.g., KLD vs. MAE). Third, when the number of identifiers is held constant, selection logic determines which fidelity dimension benefits most: The instrument-based configuration best preserves distributional shape, the data-driven configuration best recovers between-group structure and identifier–outcome associations, and the theory-driven configuration is generally intermediate. In short, no single segmentation configuration dominates all evaluative dimensions; gains in one dimension can come with weaker performance in another. Table~\ref{tab:fidelity_by_question} summarizes these patterns.

\begin{table}[htbp]
\centering
\caption{Summary of simulation performance}
\label{tab:fidelity_by_question}
\begin{tabularx}{\textwidth}{@{}>{\raggedright\arraybackslash}p{0.17\textwidth}>{\centering\arraybackslash}X>{\centering\arraybackslash}X>{\centering\arraybackslash}X@{}}
\toprule
& \textbf{Distributional fidelity} & \textbf{Structural fidelity} & \textbf{Predictive fidelity} \\
\midrule
\textbf{Granularity}     & More $\neq$ better        & More $\neq$ better        & More $\neq$ better        \\
\textbf{Parsimony}       & Mixed (metric-dependent) & Fewer generally better   & Fewer generally better \\
\textbf{Identifier selection logic} & Instrument-based best           & Data-driven and instrument-based best          & Data-driven best          \\
\bottomrule
\end{tabularx}
\end{table}

\section{Discussion}

This study demonstrates that restoring heterogeneity in LLM-based social simulation is both necessary and feasible but only when audience segmentation is designed deliberately. Moving beyond surface-level representational accuracy requires a population-thinking perspective and a robust method for operationalizing social variation. Through systematic experimentation and rigorous evaluation, our results suggest that audience segmentation can play this role by embedding individual- and subgroup-level heterogeneity into persona conditioning in a systematic and testable way.

Three key conclusions arise. First, increasing identifier granularity does not produce consistent improvements: Moving from single-dimensional to moderately enriched conditioning can improve simulation performance, but further expansion does not reliably help and may impair structural and predictive fidelity. Second, there is a clear parsimony–comprehensiveness trade-off: Additional identifiers often bring diminishing returns, and parsimonious segmentation configurations frequently match or outperform more comprehensive ones, especially on structural and predictive fidelity. Third, identifier selection logic determines which fidelity dimension benefits the most. Specifically, instrument-based identifier selection best preserves distributional fidelity, whereas data-driven logic better recovers between-group structure and identifier—outcome associations; theory-driven selection is generally intermediate. These findings shift the key focus from “how many identifiers” to “which identifiers, and why.” In short, simulation performance depends less on identifier quantity than on the theoretical or empirical relevance of the identifiers used. This echoes longstanding principles from audience segmentation and population-thinking traditions \parencite{Slater1995AmericanHealthstyles,Xie2013PopulationHeterogeneity}, which treat variation as meaningful social signals rather than residual noise.

These findings also help make sense of the mixed results in prior LLM-based social simulation studies \parencite{Argyle2023OutOfOneMany,Bisbee2024SyntheticReplacements,Wang2025MisportrayFlatten}. Some studies reported an “average persona” tendency (variance suppression) \parencite{Wang2025MisportrayFlatten,Rossi2024ProblemsLLMData,Wu2025Boundary}, while others reported unstable or uneven subgroup differentiation \parencite{Argyle2023OutOfOneMany,Tornberg2023SimulatingSocialMedia}. A plausible explanation is that these differences arose from discrepancies in LLM architecture, the intensity and objectives of reinforcement learning from human feedback, and persona prompting design \parencite{Lyman2025BalancingAlignment,Hu2025IdentityBiases,Wang2025MisportrayFlatten}. From this perspective, audience segmentation functions as a partial corrective in both directions: It can reintroduce social structure when variance is overly suppressed, and it can impose informative constraints when LLM responses become unrealistically extreme.

Methodologically, this study advances LLM evaluation practice by proposing and using a heterogeneity-aware three-dimensional fidelity framework rather than relying on surface-level average-matching. By jointly assessing distributional, structural, and predictive fidelity, we provide a more rigorous basis for assessing simulation performance. This is important as overreliance on averages risks missing the texture and complexity of social phenomena and can conceal substantial errors in subgroup differentiation \parencite{Xie2013PopulationHeterogeneity,Wang2025MisportrayFlatten,Lyman2025BalancingAlignment}. 

Our analyses also highlight several limitations. All segmentation configurations, even the best-performing ones, show residual over-regularization; that is, simulated responses remain more orderly than observed human data, and subgroup differences are still compressed or restructured. Therefore, while audience segmentation is a substantial improvement, it is not a panacea. Further progress likely requires  variance-preserving advances in model training, alignment, and calibration \parencite{Lyman2025BalancingAlignment,Wang2025MisportrayFlatten}. 

Looking forward, we recommend three priorities for further research. First, segmentation should be treated not as an afterthought or mere technicality, but as a central experimental factor that is methodologically deliberate and informed by theory rather than simply “added on” for realism \parencite{Slater1996TheoryMethod,Dibb2009ImplementationRules,Hine2014AudienceSegmentation}. Careful documentation and justification of segmentation logic will facilitate cumulative progress in the field. Second, the explicit use of heterogeneity-aware metrics should become the new gold standard for simulation evaluation. Only by quantifying and reporting the preservation (or masking) of variation can we meaningfully improve LLMs and prompt design. Third, cross-LLM comparisons should be a research priority, and attention should be paid to how different LLMs encode (or erase) social diversity \parencite{Hu2025IdentityBiases,Wang2025MisportrayFlatten,Lyman2025BalancingAlignment}. 

In conclusion, high-fidelity LLM-based social simulation is not only an engineering challenge of data scale and model training but also a methodological challenge of representing population heterogeneity. Our findings position audience segmentation as a foundational component of this agenda and offer an empirical basis for heterogeneity-centered simulation design.

\printbibliography[title={References}]

\end{document}